\documentclass[twocolumn,prl]{revtex4-2}
\usepackage{titlesec}
\usepackage{subfigure}
\usepackage{verbatim}
\usepackage{amsmath, amsfonts, amssymb, bm}
\usepackage{feynmp}
\usepackage{graphicx}
\usepackage{setspace}
\usepackage{comment}
\usepackage{dsfont}
\usepackage{color}
\usepackage[pdftex,colorlinks=true,linkcolor=blue,citecolor=blue,urlcolor=blue]{hyperref}
\usepackage{float}
\usepackage{enumerate} 
\begin{document}
\title{Non-self-averaging topological Anderson insulator}
\author{Zheng-Wei Zuo}
\email{zuozw@haust.edu.cn}
\author{Jun-Chong Liu}
\affiliation{School of Physics, Henan University of Science and Technology, Luoyang 471023, China}
\date{\today}
\begin{abstract}
Current research on the disordered topological quantum phases primarily focuses on the uncorrelated and short-range correlated disorder regime. These topological Anderson insulators are typically self-averaging. However, topological quantum systems with long-range correlated disorder have received limited attention due to the absence of tractable analytical methods. In fact, the long-range correlated disorder introduces more complex effects on topological quantum states. Here, we demonstrate that the long-range correlated disorder could induce anomalously statistical feature where the topological Anderson states become non-self-averaging, and the phase diagram is strongly dependent on individual disorder configurations. We term this statistical phase the non-self-averaging topological Anderson insulator. The non-self-averaging property is identified by the non-vanishing finite values of the relative variance of the Lyapunov exponent in the thermodynamic limit, alongside the non-Gaussian distributions of the Lyapunov exponent. Consequently, the topological properties of a single disordered sample deviate from the ensemble average, causing a breakdown of the central limit theorem. The non-self-averaging topological Anderson insulator provides insights into the interplay between correlated disorder and topology.
\end{abstract}

\maketitle

\emph{Introduction}---The past few decades have witnessed significant advances on topological quantum states due to their exotic electronic properties\cite{Bernevig13Book, AsbothJ16Book, WenXG17RMP, BergholtzEJ21RMP, MoessnerR21Book}. The discovery of integer and fractional quantum Hall effects\cite{Klitzing80PRL, Tsui82PRL} spurred condensed matter physicists to investigate physical properties of quantum matter from the topological perspective, extending beyond Landau's  paradigm. Topological quantum matter including the topological insulators, topological superconductors, and topological metals emerge from the identification of the quantum spin Hall effect\cite{Kane05PRL2, Kane05PRL, Bernevig06PRL, Bernevig06SCI}. Tens of thousands of topological quantum materials have been predicted according to the symmetry-indicator theory\cite{PoHC17NTC, KruthoffJ17PRX,  WatanabeH18SA}, topological quantum chemistry\cite{BradlynB17NT,ElcoroL21NTC} and the spin group theory\cite{XiaoZY24PRX, ChenXB24PRX, JiangY24PRX, ChenXB25NT}. These topological materials with promising implications are now attracting considerable interest from both fundamental and applied research perspectives.

The interplay between topology and disorder is a crucial subject for understanding of topological quantum states. A  key feature of these states is the robustness of these topological properties against the weak disorder. The disorder could drive the quantum systems from the trivial states into the topological Anderson states\cite{LiJ09PRL, GrothCW09PRL, JiangH09PRB, GuoHM10PRL, XingYX11PRB, StuetzerS18NT, LiuGG20PRL, CuiXH22PRL, ZhangJ22PRB, ChenR23PRB, LoioH24PRB, PadhanvA24PRB}. Various types of topological Anderson states, dependent on the specific nature of the disorder, have been proposed and investigated\cite{FuL12PRL, Ringel12PRB, MongRSK12PRL, FulgaIC14PRB, ChaouAY25PRB, ChenX25PRL, MaRC23PRX, LeeYL25QM, ZhangJH22arXiv, TaoYL23SPP, Andrea23SPP, MaRC25PRX, ChengXY23PRB, XiaoZY23PRL, ZuoZW24PRB, ZhaoPW24PRL,NeehusA25PRL, FavataB25PRL, LinJR26PRB, JiRJ25CP, ZhangGQ25CP, GrindallC25PRL, QinYL25arXiv, ZhangDW26arXiv}. For example, when disorders locally break the symmetries that are restored on disorder averaging, the average symmetry-protected topological phases\cite{FuL12PRL, Ringel12PRB, MongRSK12PRL, FulgaIC14PRB, ChaouAY25PRB, ChenX25PRL, MaRC23PRX, LeeYL25QM, ZhangJH22arXiv, TaoYL23SPP,  Andrea23SPP, MaRC25PRX} can emerge. The cooperation of Anderson disorder and structural disorder could induce the topological Anderson amorphous insulator states\cite{ChengXY23PRB}. When the disorder becomes correlated, topological Anderson states with Bloch bulk states appear\cite{ZuoZW24PRB}. The structural disorder can also drive trivial amorphous systems into topological amorphous states\cite{AgarwalaA17PRL, ManshaS17PRB, MitchellNP18NTP, PoyhonenK18NTC, YangYB19PRL, YangB19PRB, CostaM19NanoLett, ChernGW19EPL,HuangHQ20PRB,  MarsalT20PNAS, MukatiP20PRB, SahlbergI20PRR, IvakiMN20PRR, ZhouPH20LSA, CorbaeP21PRB, WangJH21PRL, LiK21PRL,SpringH21SPP, WangCT22PRL, ZhangZ23SA, MarsalQ23PRB, Munoz-SegoviaD23PRR, PengT24PRB, RegisV24PRB, PengT25PRB}. Particularly, the topological band insulators without the translational symmetry are identified in the amorphous systems\cite{WangS25PRB}.

Most studies of the disordered topological quantum phases have focused on the uncorrelated and short-range correlated disorder. The effect of the long-range correlated disorder on topological properties of the disordered quantum topological states remains an open problem, and is the subject of the present paper. The Su-Schrieffer-Heeger (SSH) model is frequently cited as an archetypal example of how topological properties can arise from symmetry protection. The disordered generalized and quasi-periodic SSH chains\cite{MondragonShem14PRL, AltlandA14PRL2,MeierEJ18SCI, HsuHC20PRB, LinL21PRB, ZuoZW22PRA, LiuSH22PLA, LuZP22AP, ShapirDS22PRA, MandaBM23PRB, ZhangH23PRB, RenMN24PRL, ZuoZW24PRB, MirandaDA24PRB, CinnirellaEG24PRB, GuY24CPB, LuZP25FP, WangXM25PRA, NairPS25PRB, SoneK25NTC, ZhangGQ26SC} have been intensively investigated. For simplicity, we consider the SSH model with long-range correlated disordered hopping as an illustrative example. We demonstrate that the long-range correlation of the disorder leads to the non-self-averaging behavior in topological Anderson insulator states, termed non-self-averaging topological Anderson insulator. The non-self-averaging property is evidenced by the finite relative variance of the Lyapunov exponent in the thermodynamic limit and the non-Gaussian distributions of the Lyapunov exponent. The phase diagram of the long-range correlated disordered SSH model is heavily dependent on individual disorder configuration due to the non-self-averaging property. In other words, the topological properties observed in a single disordered sample do not agree with the average for the whole ensemble of the disordered samples.

\emph{The long-range correlated disordered SSH chain}--- Now, to illustrate the non-self-averaging properties of the topological Anderson insulator phases, we consider the SSH chain with long-range correlated disordered hopping terms. The Hamiltonian is given by
\begin{equation}
H=\sum_{j=1}^{N}\left(  t_1+W\varepsilon_{j}\right)  c_{j,B}^{\dagger}c_{j,A}+\sum_{j=1}^{N-1}t_2 c_{j+1,A}^{\dagger}c_{j,B}+H.c.,\label{Hamiltonian}
\end{equation}
where $t_1$ and $t_2$ are the intra-cell hopping and inter-cell hopping amplitudes, respectively. The parameter $W$ represents the disorder strength, and $N$ is the number of unit cell. The $\varepsilon_{j}$ is independently randomly long-range correlated disorder with a power-law spectral density $S(k)\sim1/k^{\alpha}$\cite{deMouraF98PRL, RussS01PRB, CheraghchiH05PRB, PetersenGM13PRB, WeiXB24PRB}, and is defined as
\begin{align}
\varepsilon_{j}  =\sqrt{\frac{2\pi}{N}}\sum_{k=1}^{N/2}\left(  \frac{2\pi k}{N}\right)^{-\frac{\alpha}{2}}\cos\left(  \frac{2\pi jk}{N}+\phi_{k}\right),\label{disorder}
\end{align}
where $\phi_{k}$ are $N/2$ independent random phases uniformly distributed in the interval $\left[  0,2\pi\right] $. The parameter $\alpha$ quantifies the range of correlation. In the limit of $\alpha=0$, the long-range correlated disorder reduces to the uniform distribution case. Here, we would normalize the disorder sequence to have a mean $\left\langle \varepsilon_{j}\right\rangle =0$ and a variance$\Delta\varepsilon=\sqrt{\left\langle \varepsilon_{j}^{2}\right\rangle-\left\langle \varepsilon_{j}\right\rangle ^2}=1$. Throughout this paper, we will set inter-cell hopping $t_2=1$.

Although this disordered SSH chain is disordered, it preserves the chiral (sublattice) symmetry.  As demonstrated in Ref.\cite{MondragonShem14PRL, ZuoZW24PRB, LonghiS20OL}, the zero points of the Lyapunov exponent (which represents the inverse localization length) of the topological zero-energy modes correspond to the phase transition of the topological invariants, such as the winding number\cite{MondragonShem14PRL} and topological quantum number $\mathcal{Q}$ \cite{FulgaIC11PRB}, for the 1D chiral disordered system. Thus, we use them to identify the topological properties and topological phase transitions of the disordered SSH chain at half-filling.

The topological quantum number $\mathcal{Q}$ for this long-range correlated disordered SSH model could be expressed as
\begin{equation}
\mathcal{Q}=\frac{1}{2}\left(1-\operatorname{sgn} \left[\prod_j (t_1+W\varepsilon_{j})^2-1\right]\right).
\end{equation}
The topological quantum number $\mathcal{Q}=1$ (0) indicates a topological (trivial) insulator phase. The Lyapunov exponent of the topological zero-energy modes is given by
\begin{equation}
\gamma=\left\vert\lim_{N\rightarrow\infty}\frac{1}{N}{\displaystyle\sum\limits_{j=1}^{N}} \ln\left\vert t_{1}+W\varepsilon_{j}\right\vert \right\vert.
\end{equation}

For the disordered systems, a lack of self-averaging—or a non-self-averaging property—is exhibited when the relative variance of a quantity remains finite in the thermodynamic limit \cite{WisemanS95PRE, AharonyA96PRL, PazmandiF97PRL, WisemanS98PRE, WisemanS98PRL, ParisiG02PRL, FishJM10PRL, KrugerF11PRB, DuanY21PRL, SolorzanoA21PRR, DuthieA22PRB, AbrahamsE10Book}. In a self-averaging system, a single large sample is sufficient to represent the ensemble; however, for a non-self-averaging system, a single disordered sample does not agree with the average over the whole ensemble of the samples. Here, we investigate the relative variance of Lyapunov exponent $\gamma$, defined as
\begin{equation}
R_{\gamma}=\frac{\left\langle\gamma^{2}\right\rangle-\left\langle\gamma\right\rangle ^{2}}{\left\langle \gamma\right\rangle ^{2}},\label{variance}
\end{equation}
where $\left\langle \gamma\right\rangle =\frac{1}{N_r}\sum_i^{N_r} \gamma_i$ represents the configuration average of Lyapunov exponent with $N_r$ denoting the number of disorder realizations. The quantity $\left\langle\gamma^{2}\right\rangle=\frac{1}{N_r}\sum_i^{N_r} \gamma_i^2$ is the second moment. When the Lyapunov exponent is self-averaging, its relative variance decreases with increasing the system size, approaching zero in the thermodynamic limit and yielding agreement between the result for a single sample and the ensemble average. Consequently, this disordered SSH chain is in self-averaging topological (trivial) Anderson insulator phase. Conversely, if the relative variance of Lyapunov exponent approaches a non-vanishing finite value in the thermodynamic limit, it indicates that the disordered SSH chain is in a non-self-averaging topological (trivial) Anderson insulating phase. This results in deviations of the topological properties of a single disordered sample from the ensemble average, signifying a breakdown of the central limit theorem.

\begin{figure}[tbp]
\centering 
\includegraphics[width=0.5\textwidth]{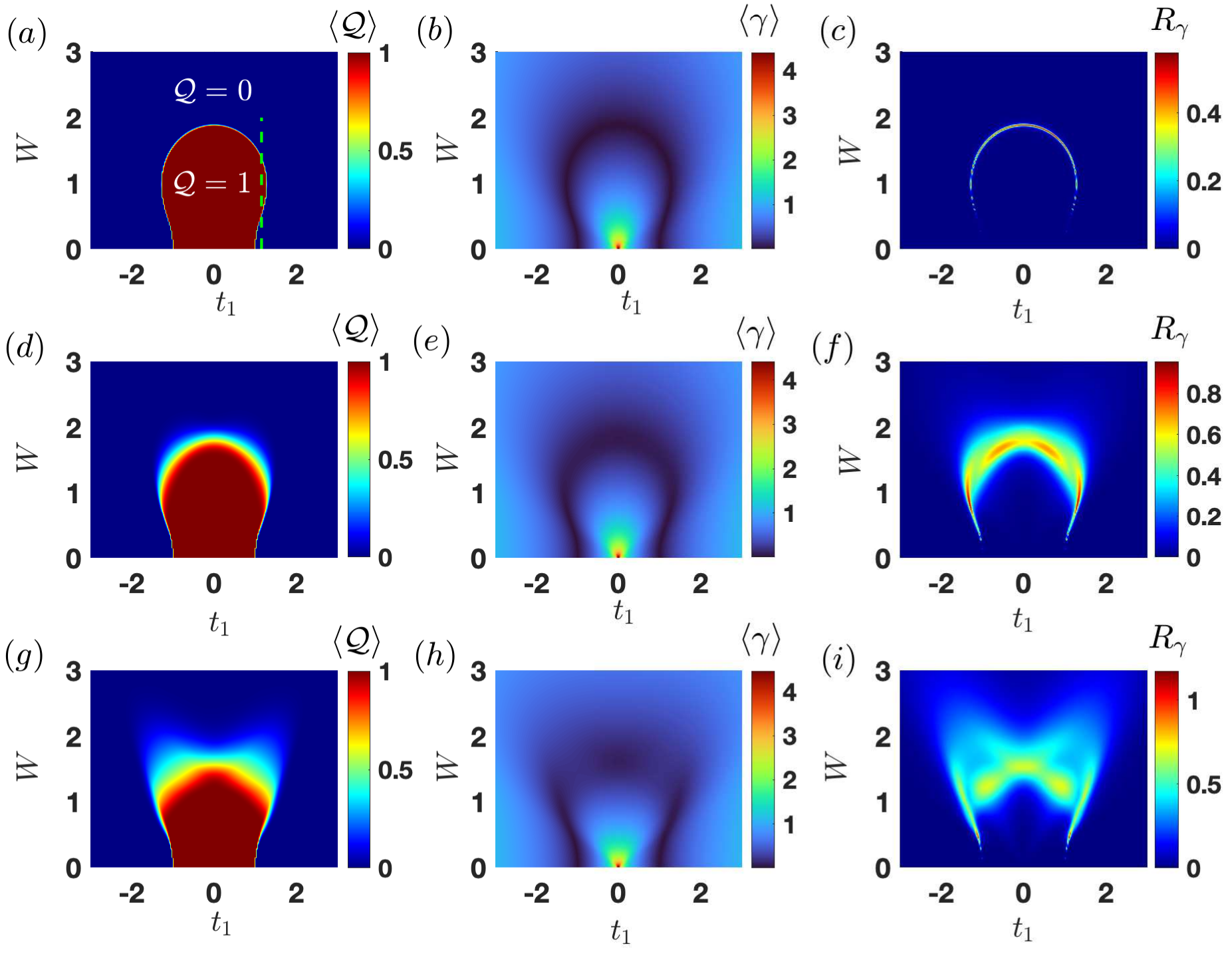}
\caption{The average of the topological quantum number $\left\langle\mathcal{Q}\right\rangle$, and the average $\left\langle\gamma\right\rangle$ and relative variance $R_{\gamma}$ of the Lyapunov exponent with the system size of $N=5\times10^4$, under $N_r=10^5$ disorder realizations. Panels $(a, d, g)$ display the average of the topological quantum number for the correlation parameters $\alpha=0$, 1.5, and 2.5, respectively. Panels $(b, e, h)$ display the average of the Lyapunov exponent for the same parameters, and panels $(c, f, i)$ show the corresponding relative variance of the Lyapunov exponent.}
\label{FigPhaseDiagram}
\end{figure}

In the following, we numerically investigate the topological properties of this long-range correlated disordered SSH chain at half-filling. Figure \ref{FigPhaseDiagram} shows the configuration average of the topological quantum number $\left\langle\mathcal{Q}\right\rangle$ (defined as $\left\langle \mathcal{Q}\right\rangle =\frac{1}{N_r}\sum_i^{N_r} \mathcal{Q}_i$), and the configuration average $\left\langle\gamma\right\rangle$ and relative variance $R_{\gamma}$ of the Lyapunov exponent, calculated using a system size $N=5\times10^4$ and $N_r=10^5$ disorder realizations. Results are presented for the correlation parameter $\alpha=0$ (Fig.\ref{FigPhaseDiagram}$[a, b, c]$), $\alpha=1.5$(Fig.\ref{FigPhaseDiagram}$[d, e, f]$), and $\alpha=2.5$(Fig.\ref{FigPhaseDiagram}$[g, h, i]$), respectively. The system highlighted in red region in Fig.\ref{FigPhaseDiagram} $(a)$ indicates the topological Anderson insulator phase. Sharp boundaries denote the topological phase transitions. For a fixed intra-cell hopping $t_1=1.15$ (indicated by green dashed line in Fig.\ref{FigPhaseDiagram}$[a]$), as the disorder strength $W$ increases gradually, the system changes from the trivial phase to the topological Anderson insulator phase. When the disorder becomes even stronger, the system enters the trivial phase again. As shown in Fig.\ref{FigPhaseDiagram}$(b)$, one can see that the averages of the Lyapunov exponent vanish at these transition points, signifying the self-averaging transition. Thus, the zero values of the Lyapunov exponent correspond to the topological phase transitions. These behaviors have been identified previously\cite{MondragonShem14PRL, ShapirDS22PRA}. The relative variances $R_{\gamma}$ (except the topological phase transition points, where the sample-to-sample fluctuations are large\cite{FavataB25PRL}) of the Lyapunov exponent approach zero as a consequence of the central-limit theorem (Fig. \ref{FigPhaseDiagram}[c]), indicating the disordered SSH chain is in self-averaging topological Anderson ($\mathcal{Q}=1$ region) or trivial insulator phases. Notably, the relative variances exhibit large maxima at topological phase transitions and this sharp crossover behavior could also be used to characterize the topological phase transitions in the self-averaging regime. As shown previously\cite{MeierEJ18SCI, HsuHC20PRB, YangYB21PRB, RenMN24PRL}, at small finite $N$, the topological invariant after statistical averaging is smooth in topological Anderson insulators. The smoothed behavior would become sharper in the large systems (see Fig.\ref{FigPhaseDiagram}$[a]$ and Refs.\cite{MondragonShem14PRL, LuZP22AP, ZuoZW24PRB}) and the self-averaging property would appear in the thermodynamic limit.

\begin{figure}[tbp]
\centering 
\includegraphics[width=0.49\textwidth]{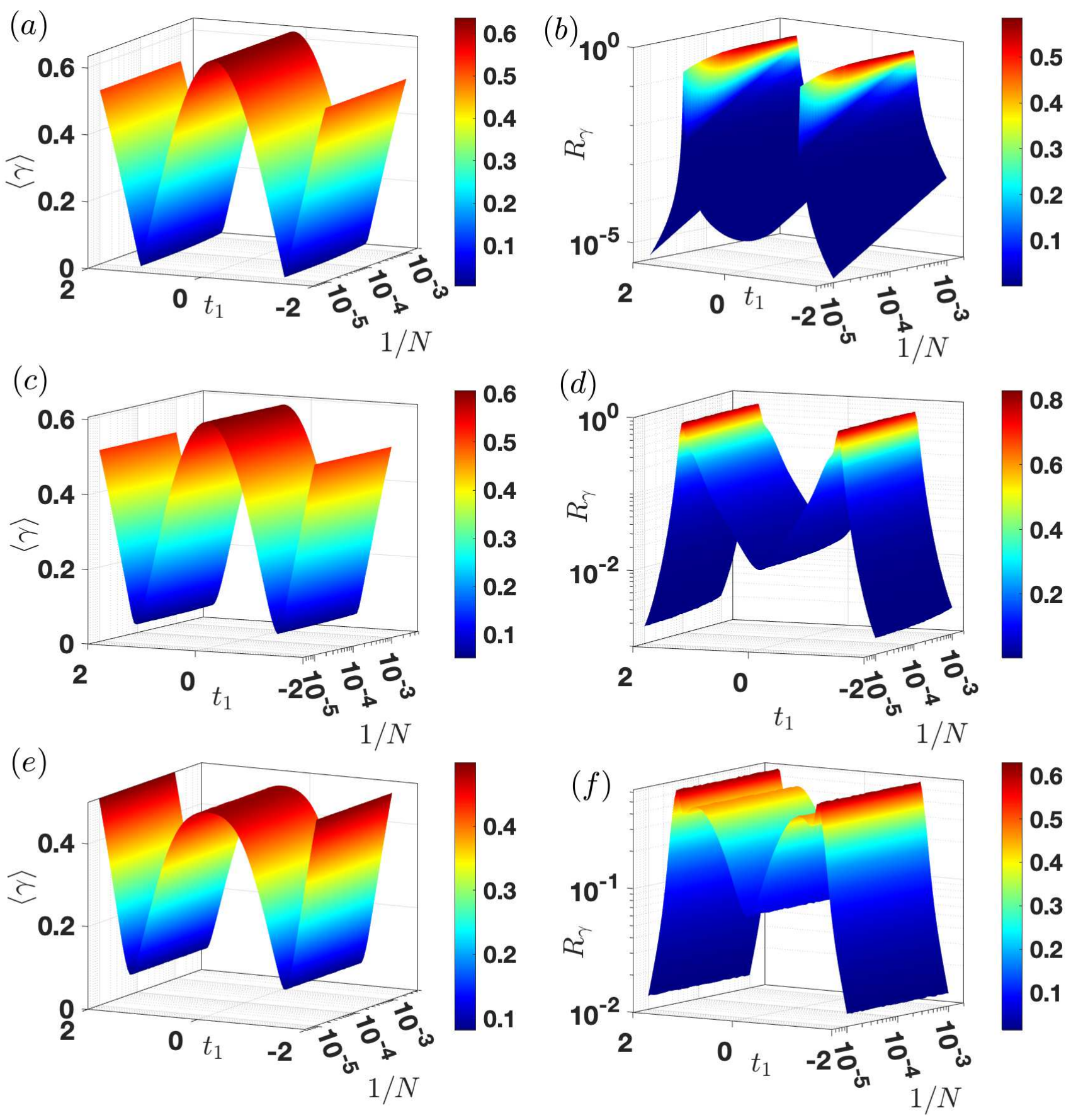}
\caption{The finite size scaling for the average and relative variance of Lyapunov exponent with the disorder strength $W=1$ under $N_r=5\times10^4$ disorder realizations. Panels $(a, c, e)$ exhibit the average of Lyapunov exponent for parameters $\alpha=0$, $1.5$ and $2.5$ cases, respectively. while panels  $(b, d, f)$ show the corresponding relative variance of Lyapunov exponent.}
\label{FigNonSelfAverage}
\end{figure}

We now investigate other $\alpha$ values, which exhibit distinct features. From Fig.\ref{FigPhaseDiagram} $(d, e)$ [$\alpha=1.5$] and Fig.\ref{FigPhaseDiagram} $(g, h)$ [$\alpha=2.5$], one can see that instead of the shape changes around topological quantum phase transitions, the crossover behaviors of topological quantum number $\mathcal{Q}$ and the Lyapunov exponent become smooth. Consequently, the topological phase transition regions between the topological Anderson and trivial insulator phases become smoothed after the ensemble averaging. Furthermore, Fig.\ref{FigPhaseDiagram} (f) [$\alpha=1.5$] and (i) [$\alpha=2.5$] demonstrate a non-vanishing finite relative variance of the Lyapunov exponent within the topological phase transition smooth regions. These anomalous behaviors indicate a breakdown of the self-averaging property, rendering the Lyapunov exponent non-self-averaging. The disordered topological SSH chain is thus in the non-self-averaging topological Anderson insulator phase. As the correlation parameter $\alpha$ further increases, our numerical calculations show that more smooth crossover behaviors appear at the original topological phase transition regions ($\alpha=0$ case). The topological phase transitions after the ensemble averaging have smooth crossover behaviors, indicating the system is in a non-self-averaging phase. In the following, we demonstrate that in the thermodynamic limit, the relative variances of the Lyapunov exponent remain finite when the system becomes non-self-averaging [see Fig.\ref{FigNonSelfAverage} $(d, f)$ for detail]. 

To investigate the fluctuations of the Lyapunov exponent in the thermodynamic limit, we plot the average Lyapunov exponent in Fig.\ref{FigNonSelfAverage} (a, c, e) and the relative variance of Lyapunov exponent as a function of system size $N$ for correlation parameter $\alpha=0$, $1.5$, and $2.5$ cases with the disorder strength $W=1$, $N_r=5\times10^4$ disorder realizations. For the $\alpha=0$ case (Fig.\ref{FigNonSelfAverage}[a]), one can see that the average Lyapunov exponent remains constant except at topological phase transitions as the system size increases. At the topological phase transition points, the Lyapunov exponent approach zeros in the thermodynamic limit.  Correspondingly, the relative variances (Fig.\ref{FigNonSelfAverage}[b]) of Lyapunov exponent approach zeros except at the topological phase transitions in the thermodynamic limit, which clearly indicates the self-averaging property. At the topological phase transition points, the relative variances retain a fixed constant value. 

In contrast, for the $\alpha=1.5$ and 2.5 cases (Fig.\ref{FigNonSelfAverage}[c, e]), the average Lyapunov exponent consistently maintains a nonzero constant value as the system size increases. So, we cannot identify the topological quantum phase transitions based on the zeros of the average Lyapunov exponent. Simultaneously, the relative variance of the Lyapunov exponent (see Fig.\ref{FigNonSelfAverage}[d, f]) remains finite in these smooth crossover regions of the topological quantum phase transitions, indicating the sample-to-sample fluctuations persist in the thermodynamic limit and excluding the influence of the finite-size effect. Thus, this long-range correlated disordered SSH chain exhibits a non-self-averaging property regardless of whether the system is in topological Anderson or trivial insulator phase. Based on the three parameters $\alpha$ cases, one can see that as the parameter $\alpha$ increases, the sample-to-sample fluctuations become stronger, and the non-self-averaging property becomes more pronounced.

\begin{figure}[tbp]
\centering 
\includegraphics[width=0.49\textwidth]{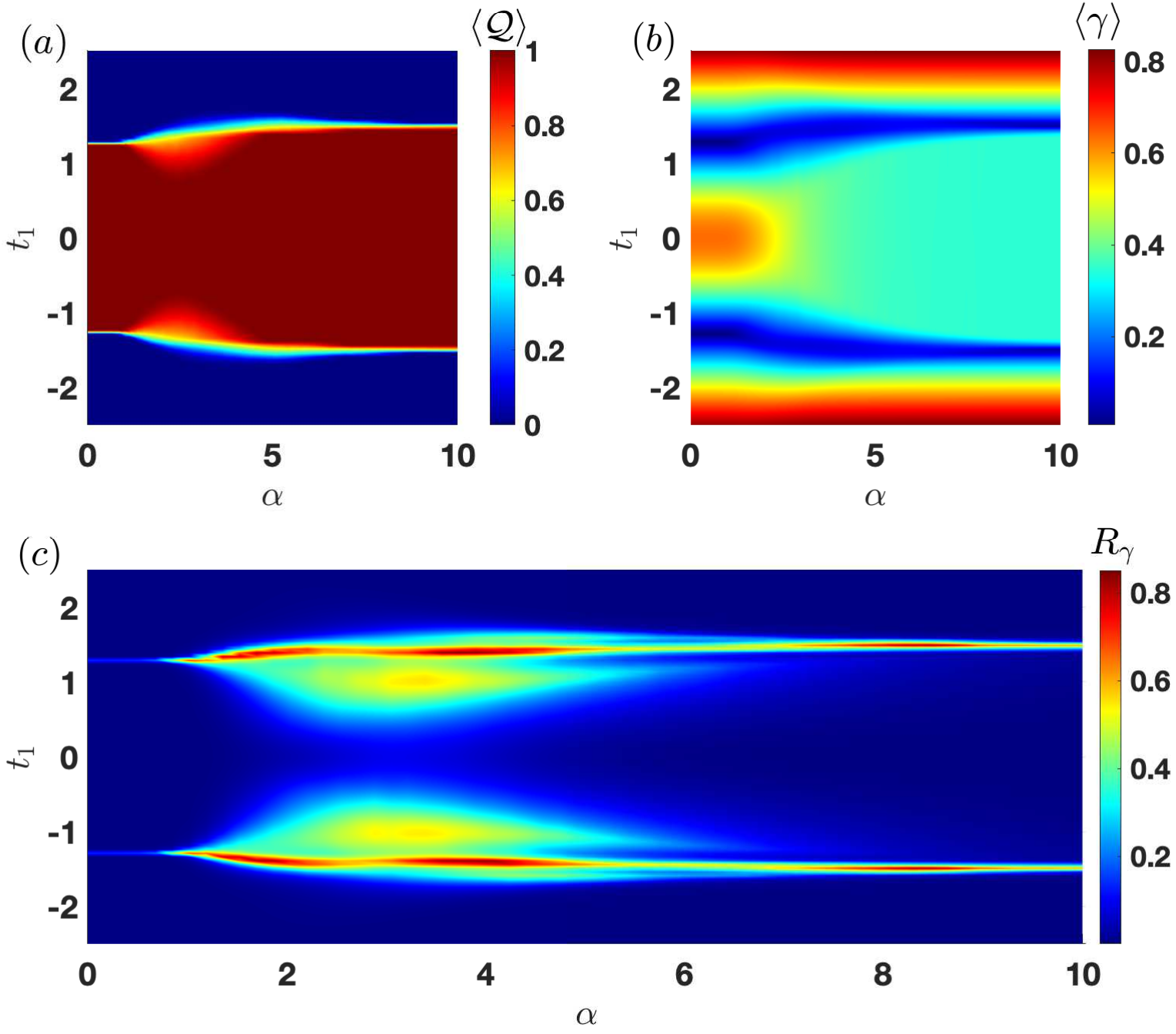}
\caption{The average topological quantum number $\left\langle\mathcal{Q}\right\rangle$, the average and relative variance of Lyapunov exponent as a function of correlation exponent $\alpha$ with disorder strength $W=1$ and system size $N=5\times10^4$, under $N_r=10^4$ disorder realizations. $(a)$ The average topological quantum number. $(b)$ The average the Lyapunov exponent. $(c)$ The relative variance of Lyapunov exponent.}
\label{FigAlpha}
\end{figure}

We next investigate the quantitative behaviors of topological quantum number and Lyapunov exponent when the parameter $\alpha$ changes. Figure \ref{FigAlpha} shows the average $\left\langle\mathcal{Q}\right\rangle$ of the topological quantum number, the average $\left\langle\gamma\right\rangle$ and relative variance $R_{\gamma}$ of Lyapunov exponent as the correlation parameter $\alpha$ increases with the disorder strength $W=1$, system size $N=5\times10^4$, under $N_r=10^4$ disorder realizations. For the small $\alpha$ ($\alpha<1$), the transitions of three physical quantities are sharp, indicating a topological quantum phase transition and preserving the self-averaging property. As the parameter $\alpha$ increases further ($1.5<\alpha<5$), one can see that the transitions become smooth and the non-self-averaging property becomes apparent. Therefore, the topological quantum phase transitions are non-self-averaging and strongly dependent on individual disorder realizations. For large parameter $\alpha$ ($\alpha>6$), the non-self-averaging property weakens while the self-averaging property strengthens.

\begin{figure}[tbp]
\centering
\includegraphics[width=0.49\textwidth]{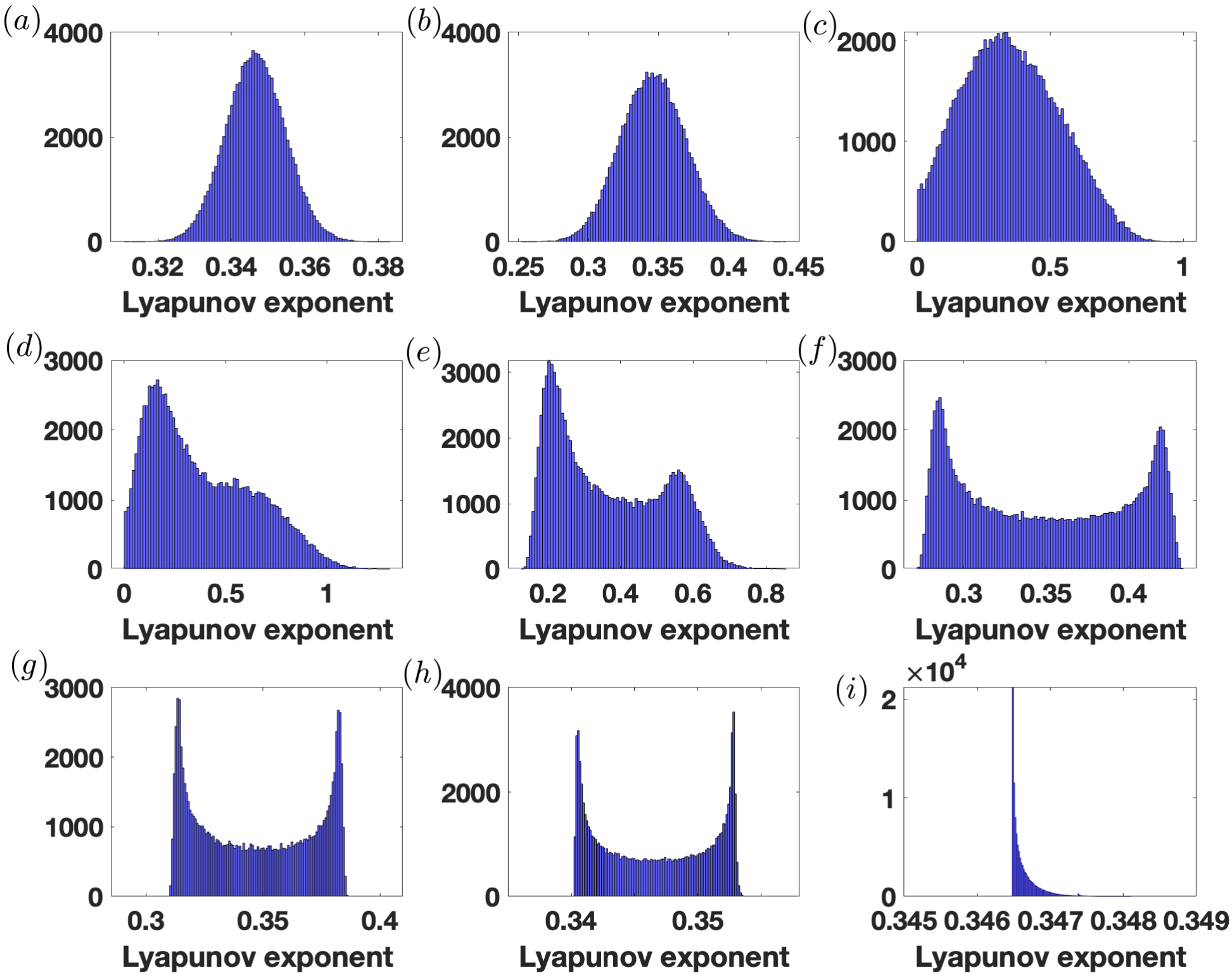}
\caption{Histogram of the Lyapunov exponent for different correlated parameters $\alpha$ with the disorder strength $W=1$,  intra-cell coupling $t_1=0.8$, and system size $N=10^4$ under $N_r=10^5$ disorder realizations. $(a)$ $\alpha=0$. $(b)$ $\alpha=1$. $(c)$ $\alpha=2$. $(d)$ $\alpha=3$. $(e)$ $\alpha=5$. $(f)$ $\alpha=8$. $(g)$ $\alpha=10$. $(h)$ $\alpha=15$. $(i)$ $\alpha=50$.}
\label{FigPDF}
\end{figure}

To visualize the extent of sample-to-sample fluctuations around the topological Anderson insulator phase transitions, we plot histograms of the Lyapunov exponent for different $\alpha$ in Fig.\ref{FigPDF} with the disorder strength $W=1$, intra-cell coupling $t_1=0.8$, and system size $N=10^4$ under $N_r=10^5$ disorder realizations. When the $\alpha$ is small ($\alpha<1$, see Fig.\ref{FigPDF}[a, b]), the histograms of the Lyapunov exponent follow a Gaussian distribution function, consistent with the central limit theorem, and preserving the self-averaging property. As the $\alpha$ increases further (Fig.\ref{FigPDF}[c-f]), the distributions become strongly asymmetric, and the Gaussian distribution breaks down, indicating a breakdown of the central limit theorem and a lack of self-averaging property. For large $\alpha$ ($\alpha>5$, see Figs.\ref{FigAlpha} and \ref{FigPDF}[e-i]), the fluctuation range of Lyapunov exponent is becoming increasingly narrow and the proportion of self-averaging property increases. According to Eq.(\ref{disorder}), one can obtain that the long-rang correlated disorder reduces to uniformly distributed disorder\cite{MondragonShem14PRL} and quasi-periodic disorder\cite{LuZP22AP, PetersenGM13PRB} for $\alpha=0$ and $\alpha\rightarrow\infty$, respectively. Thus, it is easy to understand the reentrant self-averaging behaviors as the $\alpha$ increases. For the non-self-averaging topological Anderson insulator phase, the result for a single sample does not reflect the ensemble average. Due to the asymmetric distribution of the Lyapunov exponent, the average of the Lyapunov exponent alone is inadequate to characterize the topological quantum phase transitions;  a measure of spread of the Lyapunov exponent, such as relative variance, is needed. Based on Figs.\ref{FigAlpha} and \ref{FigPDF}, one can see that as $\alpha$ increases from 0, there is no abrupt jump transition between the self-averaging and non-self-averaging regimes. Specifically, the relative variance $R_{\gamma}$ of Lyapunov exponent in Fig.\ref{FigAlpha}(c) changes smoothly with $\alpha$, lacking any discontinuous jumps. Furthermore, the histograms of the Lyapunov exponent in Fig.\ref{FigPDF} exhibit a gradual evolution—shifting continuously from a Gaussian distribution ($\alpha < 1$) to strongly asymmetric and broad distributions ($\alpha=2, 3, 5$) before narrowing again at very large $\alpha$. These features of Lyapunov exponent indicate that the degree of non-self-averaging changes continuously, characterizing a smooth crossover rather than a sharp transition.

\begin{figure}[tbp]
\centering 
\includegraphics[width=0.49\textwidth]{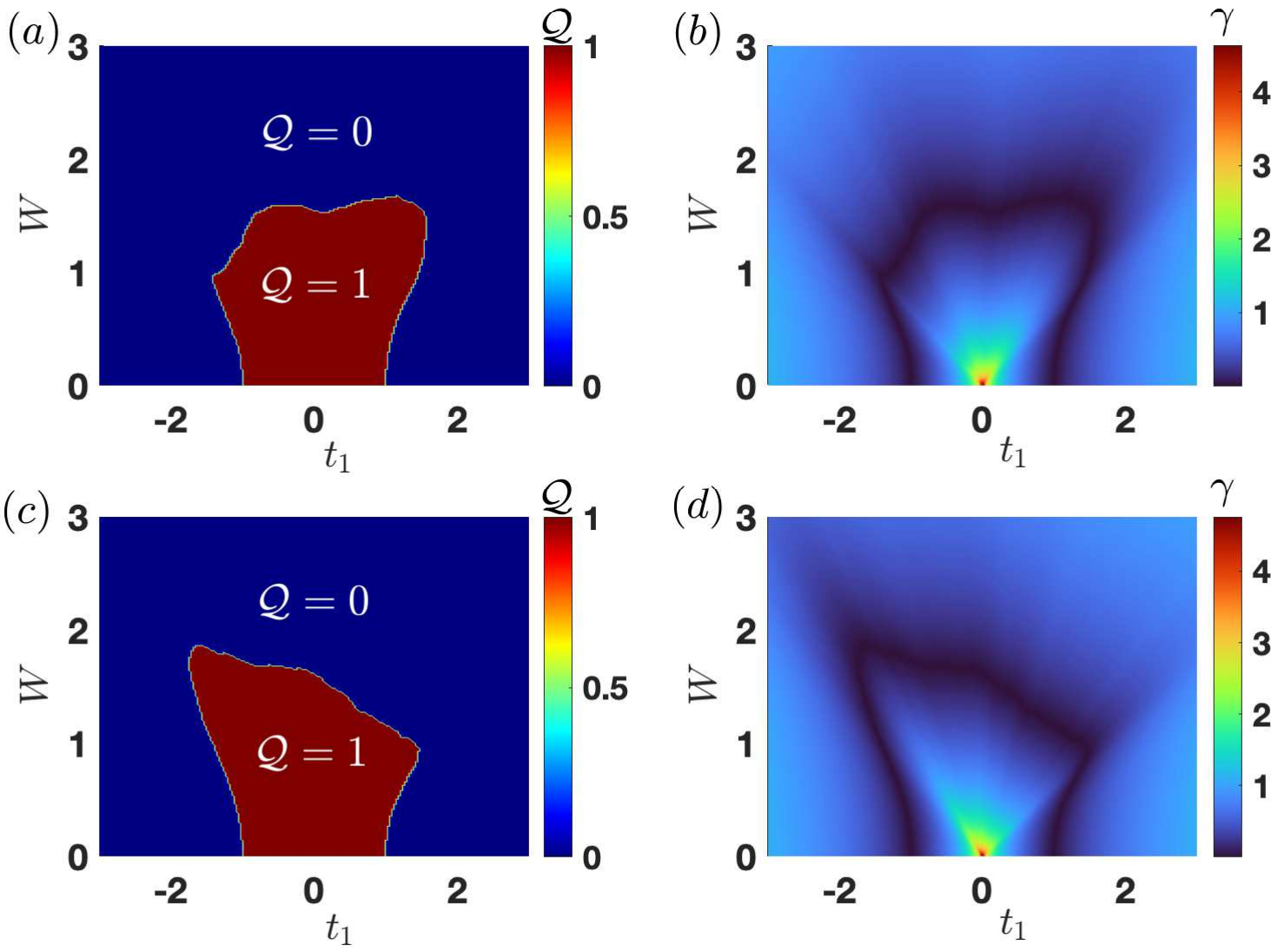}
\caption{The topological quantum number $\mathcal{Q}$ and Lyapunov exponent $\gamma$ with system size $N=10^5$ and correlation parameter $\alpha=2$ for two different disorder realizations. $(a, b)$ The topological quantum number $(a)$ and Lyapunov exponent $(b)$ for one specific disorder configuration. $(c, d)$ The topological quantum number and Lyapunov exponent for another specific disorder configuration.}
\label{FigDifferentRealizations}
\end{figure}

To vividly demonstrate the non-self-averaging property, we present the topological quantum number $\mathcal{Q}$ and Lyapunov exponent $\gamma$ for a system size of $N=10^5$ and correlation parameter $\alpha=2$ under two different configurations in Fig.\ref{FigDifferentRealizations}. For a specific realization, the zero points of the Lyapunov exponent are consistent with the changes of topological quantum number $\mathcal{Q}$, indicating topological quantum phase transitions. However, Lyapunov exponent and topological quantum number $\mathcal{Q}$ exhibit significantly different phase diagram for different disorder realizations, even though these disorder realizations share the same chiral symmetry. From the two disorder realization cases, one can see that using the average of the topological invariant and Lyapunov exponent to characterize the phase diagram of the long-range correlated disordered systems results in a smooth crossover behavior of topological phase transitions, as shown in Fig.\ref{FigPhaseDiagram}. Thus, the ensemble averages of the Lyapunov exponent and topological quantum number $\mathcal{Q}$ do not represent the behavior for a single sample, providing further evidence of the non-self-averaging property. Consequently, the values of physical quantities in systems with long-range correlated disorder cannot be solely determined from ensemble average.

Given these anomalous behaviors, we provide the qualitative analysis of the non-self-averaging property using the relative variance of the Lyapunov exponent (Eq.\ref{variance}). For large $N$, the average of Lyapunov exponent $\left\langle\gamma\right\rangle$ is determined by the average logarithm $\left\langle\ln\left\vert t_{1}+W\varepsilon_{j}\right\vert \right\rangle=\frac{1}{N}{\displaystyle\sum\limits_{j=1}^{N}} \ln\left\vert t_{1}+W\varepsilon_{j}\right\vert$ . On the other hand, for the same random Lyapunov exponent variables, its variance is equivalent to its covariance. The covariance of the Lyapunov exponent depends on the correlation of the $\left\langle\ln\left\vert t_{1}+W\varepsilon_{i}\right\vert\ln\left\vert t_{1}+W\varepsilon_{j}\right\vert\right\rangle$, which is governed by the covariance of the disordered hopping (Eq.\ref{disorder}). For this long-range correlated disorder, the covariance is given by\cite{PetersenGM13PRB}
\begin{align}
\left\langle\varepsilon_{i} \varepsilon_{j} \right\rangle=\frac{\pi}{N}\sum_{k=1}^{N/2}\left(  \frac{2\pi k}{N}\right)^{-\alpha}\cos\left(  \frac{2\pi nk}{N}\right)\label{covariance}
\end{align}
which depends solely on the distance $n=i-j$. Generally, for the long-range correlation parameter $\alpha$, the covariance of long-range correlated disorder remains finite even in the large system limit. Therefore, we can infer that non-vanishing finite relative variances of Lyapunov exponent persist in the thermodynamic limit.

In non-self-averaging phase, one may consider that the arithmetic average and arithmetic relative variance (Eq. \ref{variance}) are inadequate to characterize the long-range correlation behaviors, and propose using the geometric average and geometric relative variance instead\cite{RussS01PRB, AbrahamsE10Book}. However, our numerical calculations demonstrate that the geometric average and geometric relative variance of Lyapunov exponent also exhibit similar smooth crossover behaviors of topological quantum phase transitions, a consequence of the non-self-averaging property.

\emph{Conclusion}--- In short, using the disordered SSH model with long-range correlated disordered couplings as an example, we demonstrate the existence of the non-self-averaging topological Anderson insulators through the topological invariant and the relative variance of Lyapunov exponent. Notably, the topological invariant are not quantized under average ensemble of disorder configurations. The topological properties are non-self-averaging and the phase diagram are strongly dependent on individual disorder configurations. In essence, the topological properties for a single disordered sample does not agree with the average over the whole ensemble of the samples, indicating a breakdown of the central limit theorem. The relative variance provides an effective tool for analyzing the non-self-averaging topological states in disordered systems. Long-range correlated disorder affects the topological quantum states in more complex ways. Lastly, the non-self-averaging topological states in the long-range correlated disordered SSH model could be observed in the state-of-the-art artificial experiments, such as optical lattices\cite{MeierEJ18SCI}, photonic systems\cite{Ozawa19RMP}, phononic systems\cite{ZhuWW23RRP}, and topolectrical circuits\cite{Sahin25APLED}.

\emph{Acknowledgments.}---This work was supported by the National Natural Science Foundation of China (Grants No. 12574010 and No. 12074101) and the Natural Science Foundation of Henan (Grant No.212300410040).

 \end{document}